\documentclass[10pt,a4paper]{article}
\usepackage{amsmath,amssymb,latexsym}

\def\AFOUR{%
\setlength{\textheight}{8.5in}%
\setlength{\textwidth}{5.75in}%
\setlength{\topmargin}{-0.375in}%
\hoffset=-.5in%
\renewcommand{\baselinestretch}{1.17}%
\setlength{\parskip}{6pt plus 2pt}%
}

\AFOUR                                           

\expandafter\ifx\csname amssym.def\endcsname\relax \else\endinput\fi
\expandafter\edef\csname amssym.def\endcsname{%
       \catcode`\noexpand\@=\the\catcode`\@\space}
\catcode`\@=11
\def\undefine#1{\let#1\undefined}
\def\newsymbol#1#2#3#4#5{\let\next@\relax
 \ifnum#2=\@ne\let\next@\msafam@\else
 \ifnum#2=\tw@\let\next@\msbfam@\fi\fi
 \mathchardef#1="#3\next@#4#5}
\def\mathhexbox@#1#2#3{\relax
 \ifmmode\mathpalette{}{\m@th\mathchar"#1#2#3}%
 \else\leavevmode\hbox{$\m@th\mathchar"#1#2#3$}\fi}
\def\hexnumber@#1{\ifcase#1 0\or 1\or 2\or 3\or 4\or 5\or 6\or 7\or 8\or
 9\or A\or B\or C\or D\or E\or F\fi}

\font\tenmsa=msam10
\font\sevenmsa=msam7
\font\fivemsa=msam5
\newfam\msafam
\textfont\msafam=\tenmsa
\scriptfont\msafam=\sevenmsa
\scriptscriptfont\msafam=\fivemsa
\edef\msafam@{\hexnumber@\msafam}
\mathchardef\dabar@"0\msafam@39
\def\dashrightarrow{\mathrel{\dabar@\dabar@\mathchar"0\msafam@4B}}
\def\dashleftarrow{\mathrel{\mathchar"0\msafam@4C\dabar@\dabar@}}

\def\ulcorner{\delimiter"4\msafam@70\msafam@70 }
\def\urcorner{\delimiter"5\msafam@71\msafam@71 }
\def\llcorner{\delimiter"4\msafam@78\msafam@78 }
\def\lrcorner{\delimiter"5\msafam@79\msafam@79 }
\def\yen{{\mathhexbox@\msafam@55}}
\def\checkmark{{\mathhexbox@\msafam@58}}
\def\circledR{{\mathhexbox@\msafam@72}}
\def\maltese{{\mathhexbox@\msafam@7A}}
\def\circledS{{\mathhexbox@\msafam@73}}

\font\tenmsb=msbm10
\font\sevenmsb=msbm7
\font\fivemsb=msbm5
\newfam\msbfam
\textfont\msbfam=\tenmsb
\scriptfont\msbfam=\sevenmsb
\scriptscriptfont\msbfam=\fivemsb
\edef\msbfam@{\hexnumber@\msbfam}
\def\Bbb#1{{\fam\msbfam\relax#1}}
\def\widehat#1{\setbox\z@\hbox{$\m@th#1$}%
 \ifdim\wd\z@>\tw@ em\mathaccent"0\msbfam@5B{#1}%
 \else\mathaccent"0362{#1}\fi}
\def\widetilde#1{\setbox\z@\hbox{$\m@th#1$}%
 \ifdim\wd\z@>\tw@ em\mathaccent"0\msbfam@5D{#1}%
 \else\mathaccent"0365{#1}\fi}
\font\teneufm=eufm10
\font\seveneufm=eufm7
\font\fiveeufm=eufm5
\newfam\eufmfam
\textfont\eufmfam=\teneufm
\scriptfont\eufmfam=\seveneufm
\scriptscriptfont\eufmfam=\fiveeufm
\def\frak#1{{\fam\eufmfam\relax#1}}

\csname amssym.def\endcsname

\makeatletter
\def\section{\@startsection {section}{1}{\z@}{-3.5ex plus -1ex minus
 -.2ex}{2.3ex plus .2ex}{\large\sc}}
\def\subsection{\@startsection{subsection}{2}{\z@}{-3.25ex plus -1ex minus
 -.2ex}{1.5ex plus .2ex}{\normalsize\sc}}
\makeatother

\makeatletter
\@addtoreset{equation}{section}

\makeatother

\newcommand{\nc}{\newcommand}
\newcommand{\rnc}{\renewcommand}

\nc{\chap}[1]{{\clearpage}%
\begin{center}%
{\noindent\underline{\large\sc #1}}{\addcontentsline{toc}{section}{#1}}%
\end{center}%
{\vspace*{0.3cm}}}

\nc{\subs}[1]{{\vspace*{0.2cm}}%
{\noindent\underline{\small\sc
#1}}{\addcontentsline{toc}{subsubsection}{#1}}%
{\vspace*{0.2cm}}}

\nc{\be}{\begin{equation}}
\nc{\ee}{\end{equation}}
\nc{\bea}{\begin{eqnarray}}
\nc{\eea}{\end{eqnarray}}

\nc{\trac}[2]{{\textstyle\frac{#1}{#2}}}

\nc{\ex}[1]{\mbox{e}^{\,\textstyle#1}}

\nc{\CC}{\Bbb{C}}
\nc{\HH}{\Bbb{H}}
\nc{\PP}{\Bbb{P}}
\nc{\RR}{\Bbb{R}}
\nc{\ZZ}{\Bbb{Z}}
\nc{\II}{\Bbb{I}}
\nc{\EE}{\Bbb{E}}
\nc{\TT}{\Bbb{T}}
\nc{\DD}{\mathrm{I}\!\mathrm{D}}

\rnc{\d}{\delta}

\nc{\symx}{\circledS}
\newsymbol\smallsmile 1360
\newsymbol\smallfrown 1361
\nc{\ad}{\mathop{\mbox{ad}}\nolimits}
\nc{\tr}{\mathop{\mbox{tr}}\nolimits}
\nc{\Tr}{\mathop{\mbox{Tr}}\nolimits}
\nc{\Det}{\mathop{\mbox{Det}}\nolimits}
\rnc{\det}{\mathop{\mbox{det}}\nolimits}
\nc{\rk}{\mathop{\mbox{rk}}\nolimits}
\nc{\del}{\partial}
\nc{\diag}{\mathop{\mbox{diag}}\nolimits}
\nc{\ra}{\rightarrow}
\nc{\Ra}{\Rightarrow}
\nc{\LRa}{\Leftrightarrow}
\nc{\lra}{\leftrightarrow}
\nc{\ot}{\otimes}
\rnc{\ss}{\subset}
\nc{\nul}{\noindent\underline}
\nc{\non}{\nonumber\\}
\nc{\mat}[4]{\left(\begin{array}{cc}#1&#2\\#3&#4\end{array}\right)}
\rnc{\lg}{\frak{g}}
\nc{\G}[3]{\Gamma^{#1}_{\;{#2}{#3}}}
\nc{\nam}{\nabla_{\mu}}
\nc{\nan}{\nabla_{\nu}}
\nc{\dx}{\dot{x}}
\nc{\dxl}{\dot{x}^{\la}}
\nc{\dxm}{\dot{x}^{\mu}}
\nc{\dxn}{\dot{x}^{\nu}}
\nc{\ddx}{\ddot{x}}
\nc{\ddxm}{\ddot{x}^{\mu}}
\nc{\ddxn}{\ddot{x}^{\nu}}
\nc{\dxi}{\dot{\xi}}
\nc{\ddxi}{\ddot{\xi}}
\nc{\lsf}{\ell_s^{\mathrm{eff}}}
\nc{\lpf}{\ell_p^{\mathrm{eff}}}
\nc{\sqg}{\sqrt{g^{11}}}

\begin{document}


\vspace*{2cm}
\begin{center}
{\Large\sc Geometry of Schr\"odinger Space-Times,\\[.2in] 
Global Coordinates, and Harmonic Trapping}
\end{center}
\vspace{1cm}

\begin{center}
{\large\sc Matthias Blau, Jelle Hartong, Blaise Rollier}\\[.3cm] 
{\it Institut f\"ur Theoretische Physik\\ Universit\"at Bern\\ 
Sidlerstrasse 5 \\ 3012 Bern, Switzerland}
\end{center}

\vspace{1cm}

We study various geometrical aspects of Schr\"odinger space-times with
dynamical exponent $z>1$ and compare them with the properties of AdS
($z=1$). The Schr\"odinger metrics are singular for $1<z<2$ while the
usual Poincar\'e coordinates are incomplete for $z\geq 2$. For $z=2$
we obtain a global coordinate system and we explain the relations
among its geodesic completeness, the choice of global time, and the
harmonic trapping of non-relativistic CFTs. For $z>2$, we show that the
Schr\"odinger space-times admit no global timelike Killing vectors.


\newpage

\section{Introduction}

Recently, 
initiated by \cite{son,balamac},
there has been some interest in extending the AdS/CFT
correspondence to non-relativistic 
field theories in $d$ spatial dimensions that exhibit an anisotropic
scale invariance $(t,x^i) \ra (\lambda^z t, \lambda x^i)$ parametrised by
the dynamical critical exponent $z\geq 1$, and
corresponding to a dispersion relation of the form $\omega \sim k^z$.
While there is a plethora of non-relativistic symmetry algebras, some
of them are subalgebras of the relativistic conformal (or AdS isometry) algebra.
Systems exhibiting such a symmetry therefore potentially have bulk gravitational 
duals that can be realised as suitable deformations of AdS.
The simplest of these are the Lifshitz 
and Schr\"odinger space-times $\mathsf{Lif}_z$ \cite{lif}
and $\mathsf{Sch}_z$ \cite{son,balamac},
whose metrics in Poincar\'e-like coordinates take the form
\be
\label{metrics}
\begin{aligned}
\mathsf{Lif}_z: 
ds^2 &= -\frac{dt^2}{r^{2z}} + \frac{1}{r^2}\left(dr^2 + d\vec{x}^2 \right)\\
\mathsf{Sch}_z:  
ds^2 &= -\frac{dt^2}{r^{2z}} + \frac{1}{r^2}\left(-2dtd\xi + dr^2 + d\vec{x}^2\right)
\end{aligned}
\ee
where $d\vec{x}^2 = (dx^1)^2 + \ldots (dx^d)^2$. Subsequently, various
geometrical aspects of such a non-relativistic correspondence 
were investigated e.g.\ in
\cite{gold}-\cite{sakura}.\footnote{For an
updated account of these developments, and references to the CFT side
of the story, see also \cite{hartnoll}.} Nevertheless it is probably
fair to say that the holographic dictionary and
the issue of holographic renormalisation in these space-times are not yet nearly as
well understood as in the AdS case. 

In the usual AdS/CFT correspondence, while for most practical intents
and purposes it is sufficient to work in Euclidean signature (an
option not readily available for the $\mathsf{Sch}_z$ metrics)
or perhaps on the Minkowskian Poincar\'e patch, certain conceptual
issues of the correspondence are greatly clarified by formulating
the Lorentzian correspondence in global coordinates (see e.g.\
\cite{bkl,bklt,marolf}). For these reasons, and in order to highlight
the analogies respectively differences between AdS ($z=1$) and $z>1$,
it is important to gain a better understanding of the global geometry
of the Lifshitz and Schr\"odinger space-times.

For example, while $\mathsf{Lif}_z$ and $\mathsf{Sch}_z$
are geodesically complete at $r\ra 0$ for
all $z\geq 1$, for $z>1$ the detailed behaviour of geodesics near
$r=0$ differs somewhat from the AdS case ($r=0$ is ``harder to
reach''), and this may well have implications for holography and,
in particular, for an appropriate notion of ``boundary'' in this
context.

At the ``other end'' $r\ra\infty$, for all $z\geq  1$ the above
Poincar\'e-like coordinate system \eqref{metrics} is incomplete in
the sense that e.g. timelike geodesics can reach $r=\infty$ in
finite proper time. The implications of this run-away behaviour of
the geodesics, i.e.\ whether this indicates a genuine pathology of
the space-time (geodesic incompleteness, singularity) or a mere
coordinate singularity, requiring one to extend the space-time
beyond $r=\infty$, depend on the behaviour of the geometry
as $r\ra\infty$.  For example, it is of course well known
that in the $z=1$ AdS case the above Poincar\'e coordinates cover
only one-half of the complete (non-singular and maximally symmetric)
AdS space-time. On the other hand it has already been noted in
\cite{lif,hartnoll} that for all $z>1$ the Lifshitz geometries are
singular as $r\ra\infty$ in the sense of pp-curvature singularites
(infinite tidal forces) and are thus geodesically incomplete. For a discussion
of the possible implications of this for the 
$\mathsf{Lif}_z$ / CFT correspondence see \cite{hartnoll}.

The situation is somewhat more interesting for the Schr\"odinger metrics
$\mathsf{Sch}_z$. Our starting point is the observation that 
in this case qualitatively (for the precise statement see \eqref{tidals} below)
the tidal forces of causal geodesics behave as
\be
\label{tidal1}
\mathsf{Sch}_z: \text{Tidal Forces} \propto (z-1)r^{4-2z}\;\;.
\ee
In particular, while these space-times are geodesically incomplete for $1<
z < 2$, there are no infinite tidal forces not only for the AdS case $z=1$
but also for all $z \geq 2$, and there are freely falling
observers that reach $r=\infty$ in finite proper time without encountering
any singularity. One thus needs to provide them with a map and
extend the space-time beyond $r=\infty$.

In this note we will address this issue and obtain a global,
geodesically complete, coordinate system for $z=2$.  We also show
that the Schr\"odinger space-times for $z>2$ admit no global timelike
Killing vector fields, so that a global metric will necessarily be
time-dependent.

Taking our clue from global AdS, where global time corresponds to the generator
$P_0 + K_0$ of the isometry algebra ($K_0$ is a special conformal transformation),
we oberve that only the $z=2$ Schr\"odinger algebra has a potential counterpart of
this generator, namely $H+C$, where $H$ is the generator of $t$-translations (in the above
Poincar\'e-like coordinates) and $C$ is the special conformal generator of the $z=2$ algebra.
By considering the combination $H+\omega^2 C$ we are led to the metric
\be
\label{igma}
ds^2 =-\frac{dT^2}{R^4} + \frac{1}{R^2}(-2dT dV -\omega^2(R^2 + \vec{X}^2)dT^2 + dR^2 +
d\vec{X}^2)\;\;.
\ee
which has a number of remarkable properties. First of all,
this coordinate system, in which the metric simply has the form of a plane wave deformation of the 
Poincar\'e-like metric \eqref{metrics}, is indeed geodesically complete for $\omega>0$ and in this
sense provides global coordinates for the $\mathsf{Sch}_{z=2}$ space-time
(for $\omega=0$ the metric  reduces to the incomplete Poincar\'e-patch metric 
\eqref{metrics}). 
Moreover, this metric is closely related to the harmonic
trapping of non-relativstic CFTs that plays an important role in the
non-relativistic operator-state correspondence \cite{nison} and whose
holographic implementation was investigated in \cite{gold,bafu}. 
Our derivation of the above metric shows that precisely for $z=2$ 
(and for AdS $z=1$) the plane wave
deformation \eqref{igma} of the $\mathsf{Sch}_z$ 
Poincar\'e metric \eqref{metrics} that accomplishes this trapping
is just a coordinate transformation, namely the one that relates the
Poincar\'e time Hamiltonian $H=\del_t$ to the trapped Hamiltonian
$H+ \omega^2 C = \del_T$.
The geodesic completeness of this coordinate system
can be physically understood in terms of the
trapping of geodesics induced by the harmonic oscillator term
$\omega^2 (R^2 + \vec{X}^2)$ in the metric. Moreover, the  
spatial harmonic oscillator provides an IR cut-off that is the 
counterpart of 
the topological spatial IR cut-off (space is a sphere)
provided by AdS global coordinates.

To set the stage, in section 2 we briefly recall some elementary
aspects of the geometry (isometries, geodesics) of the $\mathsf{Lif}_z$
and $\mathsf{Sch}_z$ metrics in the Poincar\'e-like
coordinates \eqref{metrics}. In section 3.1, we motivate the introduction of $H+C$
as the generator of global time by analogy with AdS, and we show that 
the Schr\"odinger space-times for $z\neq 1,2$ have no global timelike Killing vectors.
We obtain the
desired coordinate transformation and the metric in global coordinates in
section 3.2, and in section 3.3 we establish the geodesic completeness
and discuss the other results mentioned above. In section
3.4 we briefly look at some related issues for pure AdS ($z=1$) and 
make some comments on the case $z>2$.
Finally, In section 4, we analyse the Klein-Gordon equation in global coordinates
and compare with the Poincar\'e-patch analysis of \cite{son,balamac}
and the Hamiltonian analysis of \cite{bafu}.

\section{Schr\"odinger and Lifshitz Space-Times in Poincar\'e Coordinates}

In this section, to motivate our investigation, and as a preparation
for the considerations of section 3, we briefly summarise some basic
facts about the geometry of the Schr\"odinger and Lifshitz space-times,
whose metrics in Poincar\'e-like coordinates (that we will henceforth
simply refer to as Poincar\'e coordinates) have been
given in \eqref{metrics}. Obviously for $z=1$ these reduce to the $(d+2)$-
(respectively $(d+3)$-) dimensional AdS Poincar\'e metric, 
and we will consider the range $z\geq 1$.

In addition to the manifest 
translational isometries in $t$ and $\vec{x}$ and spatial rotations
these space-times have the characteristic anisotropic dilatation symmetry
\be
\label{dil}
\begin{aligned}
\mathsf{Lif}_z:\; 
& (r,\vec{x},t) \ra (\lambda r, \lambda \vec{x}, \lambda^z t)\\
\mathsf{Sch}_z: \; 
& (r,\vec{x},t,\xi) \ra (\lambda r, \lambda \vec{x}, \lambda^z t, \lambda^{2-z}\xi)\;\;.
\end{aligned}
\ee
These comprise the so-called Lifshitz algebra \cite{lif}. The larger
Schr\"odinger isometry algebra of $\mathsf{Sch}_z$ contains, in addition,
Galilean boosts and null translations in $\xi$, the latter playing
the role of the central extension or mass operator of the Galilean
algebra. Moreover, for $z=2$ 
there is one extra special conformal
generator $C$ which will turn out to play an important role in the
considerations of section 3.

One can use the conserved momenta $E,\vec{P}$ (and $P_\xi$) corresponding
to the manifest $t$, $\vec{x}$ (and $\xi$) translational isometries of
the metrics \eqref{metrics} to reduce the geodesic equations to a single
radial (effective potential) equation
\be
\label{veff}
\begin{aligned}
\mathsf{Lif}_z: 
k &= \frac{\dot{r}^2}{r^2} + r^2 \vec{P}^2 - r^{2z} E^2\\
\mathsf{Sch}_z:  
k &= \frac{\dot{r}^2}{r^2} + r^2 (\vec{P}^2 - 2 E P_\xi)  + r^{4-2z} 
P_\xi^2
\end{aligned}
\ee
($k=0,\mp 1$ for null, timelike and spacelike
geodesics).  We will first compare and contrast the qualitative
behaviour of causal AdS geodesics ($z=1$) as $r\ra 0$ with that for $z>1$,
and then consider the (for our purposes more crucial) behaviour as
$r\ra\infty$.

In the AdS case $\mathsf{Lif}_{z=1}$ it follows from \eqref{veff}
that timelike geodesics require $E^2 - \vec{P}^2 \equiv M^2 >0$,
and that these have a minimal radius $r_{\mathrm{min}} = 1/M$, while
null geodesics ($\dot{r} = \pm M r^2$) can reach $r=0$ at infinite
values of the affine parameter. Since $\dot{t} = E r^2$, it also
follows that $r(t) = \pm (E/M)t$, so that lightrays can reach the
boundary $r=0$ and bounce back again to a stationary observer in
finite coordinate time $t$.

The behaviour of $\mathsf{Lif}_{z>1}$ causal geodesics is qualitatively
similar to the AdS case, with one perhaps crucial difference: namely,
timelike geodesics still have a minimal radius $r_{\mathrm{min}}>0$,
but here so do null geodesics unless $\vec{P}=0$ (since $r^2\vec{P}^2$
dominates over $r^{2z}E^2$ as $r\ra 0$ unless $\vec{P}=0$).  Thus
for $z>1$ only purely radial null geodesics reach $r=0$, and up to
a reparametrisation $r^z \ra r$ these are identical to null geodesics
in AdS${}_2$.

The $\mathsf{Sch}_{z>1}$ space-times exhibit a somewhat stronger
deviation from the AdS behaviour, since here neither timelike nor
null geodesics ever reach $r=0$. This is due to the fact that for
$z>1$ the dominant term in the effective potential is the positive
term $r^{4-2z}P_{\xi}^2$ unless $P_{\xi}=0$, and that there are
neither timelike geodesics, nor null geodesics with $\dot{r}\neq
0$, for $P_{\xi}=0$.

Now let us look at the behaviour as $r\ra\infty$.
It is easy to see that for all $z\geq  1$ the Poincar\'e
coordinate system \eqref{metrics} 
is incomplete. Indeed, it follows immediately from \eqref{veff} that
for $z\geq 1$ the leading large $r$ behaviour of null (and timelike) geodesics 
as functions of the affine parameter $\tau$ is 
\be 
\begin{aligned}
\mathsf{Lif}_z: 
 r(\tau)& \propto |\tau - \tau_0|^{-1/z}\\
\mathsf{Sch}_z:  
 r(\tau)& \propto |\tau - \tau_0|^{-1} \quad \forall z\geq 1
\end{aligned}
\ee
so that $r\ra\infty$ for $\tau \ra\tau_0$. Generically, i.e.\ unless 
some of the constants of motion are set to zero, all other coordinates also approach
infinity (for $\mathsf{Sch}_z$ at exactly the same rate as $r(\tau)$).

In order to assess the implications of this, one needs to look more
closely at the geometry of the space-time at $r\ra\infty$. The AdS case
$z=1$ is of course well understood: $r=\infty$ is only a coordinate singularity,
Poincar\'e coordinates cover only one-half of the complete AdS space-time, 
and it is possible to introduce global coordinates that cover the entire space-time.
It is also easy to see that (for any $z$) all scalar curvature
invariants are constant (and in particular finite at $r=\infty$). This is a consequence
of the homogeneity of the Lifshitz and Schr\"odinger space-times, in particular
the dilatation isometry \eqref{dil},
since any scalar curvature invariant can only be a function of $r$,
upon which dilatation-invariance implies that the invariant is
actually constant. 
However, as is well known, e.g.\ in the context
of pp-waves, all of whoses scalar curvature invariants are identically
zero, this does not by itself imply that the space-time should necessarily
be considered to be non-singular: freely falling observers may
nevertheless experience infinite tidal forces (and therefore a
physical singularity), in the form of divergent parallel propagated
orthonormal frame components of the Riemann tensor.

The explicit calculation of the tidal forces is pretty straightforward
in the case of Lifshitz metrics, and it has already been noted
in \cite{lif} that they are singular in this sense for $r\ra\infty$.
One finds that for null geodesics
(and up to a cosmological constant term for timelike geodesics)
the tidal forces are proportional to 
\be
\label{tidall}
\mathsf{Lif}_z: \text{Tidal Forces} \propto (z-1)r^{2z}\;\;.
\ee
This is in complete agreement with the result reported in \cite[foootnote
5]{hartnoll} and shows that the $\mathsf{Lif}_z$ space-times are
geodesically incomplete and singular at $r\ra\infty$ for all $z>1$.

For the Schr\"odinger metrics the corresponding calculation is slightly more
involved but
the result is also somewhat more interesting and 
qualitatively quite different from \eqref{tidall}. 
The relevant parallel propagated orthonormal frame components
$R^{(\tau)}_{\;(\alpha)(\tau)(\beta)}$ of the curvature tensor are
\be
\label{tidals}
\mathsf{Sch}_z: \text{Tidal Forces:} \left\{\begin{array}{ll}
R^{(\tau)}_{\;(i)(\tau)(j)} &= -(1+ P_\xi^2 (z-1) r(\tau)^{4-2z})\d_{ij}\\
R^{(\tau)}_{\;(\xi)(\tau)(\xi)} &= -(1+ 2P_\xi^2 z(z-1) r(\tau)^{4-2z}\sin(\tau)^2)\\
R^{(\tau)}_{\;(r)(\tau)(r)} &= -(1+ 2P_\xi^2 z(z-1) r(\tau)^{4-2z}\cos(\tau)^2)
\end{array}\right\}
\propto (z-1)r^{4-2z}
\ee
with $(\tau)$ referring to the tangent
of the timelike geodesic, i.e.\ $e_{(\tau)}^{\alpha} = \dot{x}^{\alpha}$.
While, as expected, $z=1$ is non-singular, this result, that can also
be deduced from the calculation of geodesic deviation in general Siklos
space-times in \cite{podolski} and \cite{cgr}, shows some perhaps surprising
features.

Namely, while a static (non-geodesic) observer may have have been
inclined to believe that the metric is asymptotically AdS for
any $z>1$ as $r\ra\infty$, since the $r^{-2z}dt^2$ term appears to be subleading, 
this is an illusion caused by that observer's acceleration.
Indeed \eqref{tidals} shows that there is a singularity in the form of
infinite tidal forces at $r=\infty$ for $1< z < 2$, experienced by all
timelike and null geodesics (since for these $P_{\xi}\neq 0$), and the
situation is therefore very much like that of the Lifshitz space-times
for $z>1$.  For $z\geq 2$, however, the tidal forces again remain finite
as $r\ra\infty$.\footnote{For $z>2$, the dangerous region may appear to
be $r\ra 0$, but this is deceptive since (as discussed above) timelike
and null geodesics have a non-zero minimal radius $r_{\mathrm{min}}>0$.}
\textit{Mutatis mutandis} the above result is also valid for the tidal
forces experienced by null geodesics; the first (cosmological constant)
term does not contribute in that case.  Thus for $z\geq 2$ causal
geodesics reach $r=\infty$ at finite values of the affine parameter
without encountering any singularity.

\section{Global Coordinates for $z=2$ Schr\"odinger Space-Times}

The above analysis points to the
necessity of constructing suitable coordinates that cover the space-time
region beyond $r=\infty$.  In this section we will obtain a global,
geodesically complete, coordinate system for $\mathsf{Sch}_{z=2}$ and
describe some of its properties. We also make some comments on the 
(qualitatively quite different) case $z>2$.

\subsection{Towards Global Coordinates}

We begin by recalling the situation for AdS, i.e.\ $\mathsf{Sch}_{z=1}$. In this case, 
it is well known how to construct global coordinates $(T,R,\text{angles})$ in terms of
which the $(d+3)$-dimensional AdS metric takes the form 
\be
\label{adsg}
\mathsf{AdS:}\; ds^2 = -(1+R^2)dT^2 + (1+R^2)^{-1}dR^2 + R^2 d\Omega_{d+1}^2\;\;.
\ee

The most straightforward way to find these global coordinates is to make use of the embedding of 
the unit curvature radius $(d+3)$-dimensional AdS space-time
into $\RR^{2,d+2}$ with coordinates $Z^A, A=0,\ldots, d+3$ and metric 
\be 
ds^2 = -(dZ^0)^2 + (dZ^1)^2 + \ldots + (dZ^{d+2})^2 - (dZ^{d+3})^2
\ee
as the (universal covering space of the) hyperboloid 
\be
\label{adsem}
-(Z^0)^2 + (Z^1)^2 + \ldots + (Z^{d+2})^2 - (Z^{d+3})^2 = -1\;\;.
\ee
Writing this as
\be
(Z^0)^2 + (Z^{d+3})^2 = 1 + (Z^1)^2 + \ldots + (Z^{d+2})^2 \equiv 1+R^2
\ee
suggests the parametrisation
\be
Z^0 = (1+R^2)^{1/2} \sin T\qquad Z^{d+3} = (1+R^2)^{1/2} \cos T \;\;,
\ee
which identifies $\del_T$ with the generator of rotations 
$M_{0,d+3}$ in the timelike $(Z^0,Z^{d+3})$-plane. 
This indeed gives rise on the nose to the global metric \eqref{adsg}. 

Since the $\mathsf{Sch}_z$ metric \eqref{metrics} differs from the AdS
Poincar\'e metric only by the characteristic first term $-dt^2/r^{2z}$,
when seeking global coordinates for $\mathsf{Sch}_z$, one's first thought
may perhaps be to simply employ the usual transformation from AdS Poincar\'e
coordinates
\be
\label{adsp}
ds^2 = \frac{-dt^2 + d\vec{y}^2 + dr^2}{r^2}\;\;,
\ee
to global coordinates. However, since e.g.\ the relation between global and Poincar\'e time
is 
\be
\label{adsT}
\tan T = \frac{2t}{1+ r^2 + \vec{y}^2 - t^2}\;\;,
\ee
this results in a fairly complicated metric that explicitly depends on all of the coordinates. In
particular, none of the Schr\"odinger isometries of the metric will be manifest and, regardless of
whether or not this procedure leads to a geodesically complete coordinate system for $z\geq 2$, it appears 
to provide no additional insight into the geometry of Schr\"odinger space-times. 

Another possibility is to try to find a deformation of the AdS embedding \eqref{adsem} that
breaks the conformal algebra down to its Schr\"odinger sub-algebra.
Unfortunately, it is not hard to see that such an embedding of
$\mathsf{Sch}_z$ into one dimension higher does not exist: any
hypersurface invariant under the Schr\"odinger algebra turns out to be
automatically invariant under the entire conformal algebra, and leads
to the standard AdS hyperboloid.

However, there is yet another aspect of the AdS construction that
does turn out to generalise to the Schr\"odinger case, and does
provide global coordinates, but only for $z=2$. Namely, 
under the usual
identification of the generators $M_{AB}$ of $\frak{so}(d+2,2)$ with the generators 
$(P_{\mu},K_{\mu},M_{\mu\nu},D)$ of the relativistic conformal algebra $\frak{conf}(d+1,1)$, such that
e.g.\ 
\be
P_0 = \del_t \quad,\quad K_0 = t(r \del_{r} + y^{i}\del_{y^i})+ \trac{1}{2}(t^2 + r^2 + \vec{y}^2)\del_t 
\ee 
in standard Poincar\'e coordinates \eqref{adsp},
the definition of AdS global time is equivalent to the identification 
\be
\label{adstpk}
\del_T = P_0 + K_0\;\;.
\ee
Thus global AdS time ``diagonalises'' the modified Hamiltonian operator $P_0 +
K_0$. In the Schr\"odinger algebra, the role of the
Hamiltonian is played by the lightcone Hamiltonian $H \equiv P_+$,
and the Poincar\'e coordinates \eqref{metrics} are such that
this Hamiltonian is diagonalised, $H = \del_t$. Now, generically the
Schr\"odinger algebra does not possess any counterpart of the special
conformal generators $K_{\mu}$. Precisely for $z=2$, however (for which
the Schr\"odinger algebra can be characterised as the subalgebra of 
$\frak{conf}(d+1,1)$ that commutes with the lightcone momentum $P_{-}$),
there is one extra special conformal generator, namely $C\equiv K_-$.
In Poincar\'e coordinates $C$ takes the form
\be
\label{C}
C = t(t\del_t + r\del_r + x^i\del_{x^i}) + \trac{1}{2}(r^2 + \vec{x}^2)\del_{\xi} 
\ee
with 
\be
{}[H,C] = D\qquad [D,C]=2C \qquad [D,H] = -2H\;\;,
\ee
where
\be
D=2t\del_t + r\del_r + x^i\del_{x^i}
\ee
is the generator of dilatations \eqref{dil} for $z=2$. Thus for $z=2$, there is a natural candidate counterpart
of the AdS global Hamiltonian $P_0 + K_0$, namely
\be
\label{hg}
P_+ + K_- = H + C\;\;.
\ee
As a first check on this we can calculate the norm of this Killing vector in the
Poincar\'e-coordinates of \eqref{metrics}, 
\be
||H+C||^2 = -1-\frac{\vec x^2}{r^2}-\frac{(1+t^2)^2}{r^4} \leq -1\;\;.
\ee
Here the constant term $-1$ arises from the cross-term between $\del_t$ and the
$r^2$-term in $C$ \eqref{C}.
Thus unlike $H=\del_t$, whose norm goes to zero as $r\ra\infty$, this Killing vector
is everywhere timelike in the Poincar\'e patch and thus has a chance of providing a 
well-defined notion of time also beyond the Poincar\'e patch.
We will show below that diagonalising this generator of the isometry algebra
indeed leads to global time (and other global coordinates) for $z=2$.

Before turning to that, let us briefly look at the situation for
$z\neq 1,2$.  In that case, $C$ is absent but one could e.g.\
consider a linear combination $H + aD + bP_-$, $P_-=\del_\xi$, of
the Killing vectors that are invariant under spatial rotations.
Such Killing vectors necessarily become spacelike somewhere inside the
Poincar\'e patch if $a\neq 0$, while the norm of $H + bP_-$ still
goes to zero as $r\ra\infty$ (in any case, 
replacing $H$ by $H+ bP_-$ just
amounts to passing from $(t,\xi)$ to some linear combinations of $t$ and
$\xi$). Including the remaining Killing vectors (rotations, translations,
boosts) in this analysis does not improve the situation. We can
therefore conclude that, unlike for $z=2$,
the Schr\"odinger space-times $\mathsf{Sch}_z$
for $z>2$ have no global timelike Killing vector fields. We will
come back to this result in section 3.4.

\subsection{Global Coordinates for $z=2$}

Since the $z=2$ algebra 
has the central element $P_-=\del_\xi$, we seek new coordinates 
\be
(t,r,\vec{x},\xi) \mapsto (T,R, \vec{X},V)
\ee
in which $H+C$ \eqref{hg} and $P_-$ are simultaneously diagonal, 
\be
H+C = \del_T\quad,\quad P_- = \del_V\;\;.
\ee
This is accomplished by the coordinate transformation 
\be
\label{ct1}
\begin{aligned}
t&=\tan T\quad,\quad r=\frac{R}{\cos T}\quad,\quad \vec{x}= \frac{\vec{X}}{\cos T}\\
\xi &= V + \frac{1}{2}\left(R^2 + \vec{X}^2\right) \tan T
\end{aligned}
\ee
(chosen to also keep the metric as diagonal as possible - 
no off-diagonal terms in the new radial coordinate $R$ - see also section 3.4 for 
further comments on this transformation), and in these coordinates
the metric reads
\be
\label{gm1}
\begin{aligned}
\mathsf{Sch}_{z=2}:\;
ds^2 &=-\left( \frac{1}{R^4} + (1 + \frac{\vec{X}^2}{R^2})\right)dT^2 + \frac{1}{R^2}(-2dT dV + dR^2 +
d\vec{X}^2)\\
&=-\frac{dT^2}{R^4} + \frac{1}{R^2}(-2dT dV -(R^2 + \vec{X}^2)dT^2 + dR^2 +
d\vec{X}^2)\;\;.
\end{aligned}
\ee
This metric has several noteworthy properties. First of all, it is indeed
geodesically complete, i.e.\ all geodesics can be extended to infinite
values of their affine parameter. We postpone a detailed proof of this
assertion to section 3.3, but already here draw attention to the fact
that a crucial role in establishing this is played by the
harmonic oscillator potential $R^2 + \vec{X}^2$ induced by the
isotropic plane wave metric $-2dT dV -(R^2 + \vec{X}^2)dT^2 + dR^2 +
d\vec{X}^2$ in \eqref{gm1}
which replaces
the flat Poincar\'e coordinate metric $-2dtd\xi + dr^2 + d\vec{x}^2$.
In section 3.3 we will also discuss other
issues related to this term and its interpretation in terms of the harmonic
trapping \cite{nison} of non-relativistic conformal field theories.

It is perhaps quite surprising that the above transformation between
Poincar\'e and global cooordinates is so much simpler than its AdS
counterpart. For instance, instead of the AdS relation \eqref{adsT}
one has the simple relation $t=\tan T$ \eqref{ct1} between Poincar\'e
and global time for $z=2$.  The success of the simple coordinate
transformation \eqref{cta} can be traced back or attributed to the
fact (we will briefly recall in section 3.4) that it is precisely
the coordinate transformation that exhibits the conformal flatness of
isotropic plane waves.

One remarkable (and related) feature of the global metric \eqref{gm1}
is that it differs from the Poincar\'e-metric \eqref{metrics} only by a
single term, namely the plane wave harmonic oscillator frequency term
$\sim (dT)^2$.  This is in marked contrast to the global AdS metric
\eqref{adsg} which appears to bear no resemblance whatsoever to the
Poincar\'e metric. This statement can even be sharpened somewhat by
introducing a real (and without loss of generality positive) parameter
$\omega$ into the coordinate transformation \eqref{ct1}, via
\be
\label{cta}
\begin{aligned}
t&=\omega^{-1}\tan \omega T\quad,\quad r=\frac{R}{\cos \omega T}\quad,\quad \vec{x}= \frac{\vec{X}}{\cos \omega T}\\
\xi &= V + \frac{\omega }{2}\left(R^2 + \vec{X}^2\right) \tan \omega T\;\;.
\end{aligned}
\ee
In terms of these coordinates the metric now takes the form 
\be
\label{gma}
\mathsf{Sch}_{z=2}:\;
ds^2 =-\frac{dT^2}{R^4} + \frac{1}{R^2}(-2dT dV -\omega^2(R^2 + \vec{X}^2)dT^2 + dR^2 +
d\vec{X}^2)\;\;.
\ee
Thus this metric interpolates between the Poincar\'e metric for $\omega =0$
(for which \eqref{cta} obligingly reduces to the identity transformation)
and the global metric for ${\omega}=1$.  The metric is actually geodesically
complete for any ${\omega} > 0$ since
\eqref{gma} can be obtained from \eqref{gm1} by the scaling 
$(R,T,\vec{X},V)\ra (\sqrt{{\omega}}R,{\omega}T,\sqrt{{\omega}}\vec{X},V)$. This happens to
looks very much like the
$z=2$ dilatation symmetry \eqref{dil} in Poincar\'e coordinates, but acting
on global coordinates this is not an isometry but rather the transformation 
that turns \eqref{gm1} into \eqref{gma}. 

To better understand what is going on here, note that the coordinate
transformation \eqref{cta} diagonalises not $H+C$ but $H+{\omega}^2 C$,
so that it is not too surprising that one finds the Poincar\'e
metric for ${\omega}=0$ and the global metric \eqref{gm1} for ${\omega}=1$.  Thus
any non-trivial linear combination of $H$ and $C$ (with a relative
positive coefficient) gives rise to a global Hamiltonian.  

While this explains the form \eqref{gma} of the $z=2$ global metric,
this begs the
question if one cannot adopt a similar procedure in the AdS case,
diagonalising not $P_0 + K_0$ (which, as we know, gives rise to global coordinates)
but $P_0 + {\lambda}^2 K_0$ and finding a metric
depending on ${\lambda}$ that interpolates between the Poincar\'e metric
for ${\lambda}=0$ and the global
metric for ${\lambda}=1$. At first this may perhaps appear unlikely precisely because the global
metric \eqref{adsg} is so unlike the Poincar\'e metric, but nevertheless
this is indeed possible. For notational simplicity, we exhibit this
1-parameter family of interpolating metrics only in the $(3+1)$-dimensional
($d=1$) case, 
\be
\label{adsb}
\mathsf{AdS}:\; ds^2 = -({\lambda}^2 + R^2) dT^2 + ({\lambda}^2 + R^2)^{-1} dR^2 + R^2 
\left(d\Theta^2 + \cos^2(\frac{{\lambda}^2\pi}{2} - {\lambda} \Theta)d\Phi^2\right)\;\;.
\ee
The explicit coordinate
transformation, which we will not give here in detail (instead of 
\eqref{adsT} one now has
\be
\label{adsTa}
\tan {\lambda}T = \frac{2{\lambda}t}{1+ {\lambda}^2(r^2 + \vec{y}^2 - t^2)}
\ee
etc.) shows that $({\lambda}T,{\lambda}\Theta,{\lambda}\Phi)$ are standard angles. Thus,
on the one hand the above metric reduces to the global metric 
\eqref{adsg} for ${\lambda}=1$, 
while on the other hand for ${\lambda}\ra 0$ the time-coordinate
becomes non-compact and the spatial sphere decompactifies to Euclidean space, so
that one obtains the standard Poincar\'e metric \eqref{adsp} with $R=1/r$ and $\{y^i\}=\{\Theta,\Phi\}$.

\subsection{The Global Metric: Harmonic Trapping and Geodesic Completeness}

There is one aspect of the global $\mathsf{Sch}_{z=2}$ metric
constructed above that merits particular attention, and that we already alluded 
to above, namely its
relation to the harmonic trapping of non-relativistic CFTs \cite{nison}
and its geometric realisation \cite{gold,bafu}. Recall that we were led
to the metric \eqref{gm1} by analogy with the AdS case and by the
realisation that there is an essentially unique counterpart of the
global AdS Hamiltonian $P_0 + K_0$ in the $z=2$ Schr\"odinger
algebra, namely the generator $P_+ + K_- \equiv H+C$.

Non-relativistic CFT, on the other hand, provides an a priori
completely different rationale for studying the modified Hamiltonian
$H \ra H+C$, because the non-relativistic operator-state correspondence
\cite{nison} relates primary operators of the Schr\"odinger algebra
(those that commute with $C$ and Galilean boosts) with energy
eigenstates of $H+C$. Since essentially the effect of $C$ is to add
a harmonic potential to the Hamiltonian, this corresponds to putting
the system into a harmonic trap.

In \cite{bafu}, the question was investigated how this trapping could
be realised holographically via a deformation of the (Poincar\'e patch)
Schr\"odinger metric \eqref{metrics}. The deformation that was found
to accomplish a harmonic trapping both in the spatial directions of
the CFT and in the holographic radial coordinate $r$ turns out (when
specialised to $z=2$)\footnote{The emphasis in \cite{bafu} (and also in
\cite{gold}) was on $z=1$ and the attempt to find a pure AdS DLCQ dual
realisation of systems with $z=2$ Schr\"odinger symmetry.} to agree
precisely with the global metric \eqref{gma}. Our derivation of this
metric shows that precisely for $z=2$ (and for $z=1$, see section 3.4)
the required deformation of the metric that accomplishes this trapping is
actually a pure gauge deformation, namely a coordinate transformation
that relates the Poincar\'e time generator $H=\del_t$ to the trapping
(or global) time generator $H+{\omega}^2 C= \del_T$.

The plane wave deformation \eqref{gma} of the Poincar\'e metric \eqref{metrics}
achieves this trapping deformation of the Hamiltonian for the same
reason that the massless Klein-Gordon equation for a scalar field
$\Phi$ in a pp-wave metric background
\be
ds^2 = - 2dtd\xi - U(\vec{x},t) dt^2 + d\vec{x}^2
\ee
reduces
to the Schr\"odinger equation with a potential $V = mU/2$ in a sector with
fixed lightcone momentum = mass, 
\be
\Box \Phi = 0 \;,\; i\del_{\xi}\Phi = m\Phi
\quad \Ra\quad i\del_t \Phi = - \frac{1}{2m} \Delta \Phi +
\frac{m}{2}U\Phi\;\;.
\ee
For an isotropic harmonic oscillator potential, this is precisely
the plane wave metric that appears in \eqref{gma}, and we will also
encounter this trapping of the scalar field in the analysis of the
Klein-Gordon equation in the metric \eqref{gma} in the next section.

We will now show that
the completeness of the coordinate system
\eqref{cta}
is a consequence of the harmonic trapping of geodesics induced by this coordinate
transformation. To study the effect of the harmonic oscillator term
${\omega}^2(R^2 + \vec{X}^2)$
in the metric on the behaviour of geodesics, let us compare
the $z=2$ Poincar\'e radial effective potential equation \eqref{veff}
\be
\label{vp}
k = \frac{\dot{r}^2}{r^2} + r^2 (\vec{P}^2 - 2 E P_\xi)  + P_\xi^2
\ee
with the corresponding equation one obtains from the global metric
\eqref{gma}, namely 
\be
\label{vg}
k = \frac{\dot{R}^2}{R^2} + R^2 (P^2 - 2 E P_V)  + P_V^2 + {\omega}^2P_V^2 R^4\;\;.
\ee
A minor difference between 
\eqref{vp} and \eqref{vg} is the fact that the constant of
motion denoted by $P^2$ in \eqref{vg} arises not like the $\vec{P}^2$-term in
\eqref{vp} as a consequence of translation invariance (which \eqref{gma} does not
manifest), but rather as the conserved energy 
\be
P^2 \equiv \left(\frac{1}{R^2} \frac{d}{d\tau} \vec{X}\right)^2 
+ {\omega}^2 P_V^2 \vec{X}^2 
\ee
associated to the transverse harmonic
oscillator equations 
\be
\frac{1}{R^2} \frac{d}{d\tau} \left(\frac{1}{R^2} \frac{d}{d\tau} 
\vec{X}\right) = - {\omega}^2 P_V^2 \vec{X}\;\;.
\ee
The main (and crucial) difference between \eqref{vp} and \eqref{vg}, however,
lies in the last term ${\omega}^2 P_V^2 R^4$ in \eqref{vg}. This term is negligible for
$R\ra 0$, where \eqref{vg} reduces to \eqref{vp} which, as we already discussed, is
well-behaved as $r\ra 0$. On the other hand, since this term dominates the
large $R$ behaviour, it prevents any geodesic (for any $k$) from reaching $R=\infty$
(even for infinite values of the affine parameter $\tau$) unless ${\omega}=0$ (the
Poincar\'e-patch metric, which as we know is incomplete at large radius) or $P_V=0$.
When $P_V=0$, the right-hand-side of \eqref{vg} is a sum of squares, and thus only 
spacelike geodesics $k=+1$ are possible. When $P^2\neq 0$, there is again 
a maximal radius, $R_{\mathrm{max}}= 1/P$, and when $P_V=P=0$, one has $\dot{R}=\pm
R$, and thus these are the only geodesics that can reach $R=\infty$, but they only do
so for $|\tau|\ra\infty$.
The motion in the $\vec{X}$-direction is bounded by the harmonic oscillator potential,
and that in the remaining $(T,V)$-directions is determined by that of $R$ and
$\vec{X}$ and remains at finite values of the coordinates for all finite $\tau$.
This establishes that, as claimed in section 3.2, the $\mathsf{Sch}_{z=2}$
metric written in the coordinates (\ref{gm1},\ref{gma}) is geodesically
complete.

We close this section with one more remark on the significance of the trapping
exhibited by the global metric and the comparison with the global AdS metric
\eqref{adsg}. As is well known the slices of constant $R$ there have the topology
$\RR\times S^{d+1}$.
\be
\begin{aligned}
\mathsf{AdS:}\; ds^2|_{R=const} &= -(1+R^2)dT^2  + R^2 d\Omega_{d+1}^2\\
&\stackrel{R\ra \infty}{\longrightarrow} 
R^2(-dT^2 + d\Omega_{d+1}^2)
\end{aligned}
\ee
Thus the spatial part of the induced
boundary metric  has topology $S^{d+1}$, with finite volume,
and thus in particular provides a topological IR cut-off for the boundary theory.
Without committing ourselves to a particular notion
of boundary in the Schr\"odinger case, roughly speaking the dual CFT should be
considered to live on the slices of constant $R$ (the holographic coordinate) and $V$
(dual to the particle number). The metric induced on these slices can, via some
constant rescalings of the coordinates, be written as
\be
\label{rvc}
\begin{aligned}
\mathsf{Sch}_{z=2}:\;ds^2|_{R,V = const.} &\sim -(1 + {\omega}^2|\vec{X}|)^2 dT^2 + d\vec{X}^2 \\
&= -(1 +{\omega}^2\rho^2 ) dT^2 + d\rho^2 + \rho^2 d\Omega_{d-1}^2\;\;.
\end{aligned}
\ee
Thus in the Schr\"odinger case there is no topological cut-off,
but the trapping in the spatial directions $\vec{X}$ can be thought
of as providing an IR cut-off through the harmonic potential.
In particular, the induced metric \eqref{rvc} has the standard 
form of the Newtonian limit of a relativistic metric, here in a spherically
symmetric gravitational 
harmonic oscillator potential $\frac{1}{2}{\omega}^2\rho^2$. 
This Newtonian limit aspect of the metric
of course fits in well with the non-relativistic symmetries and potential dual
dynamics.

\subsection{Some Comments on $z=1$ and $z>2$}

While we have motivated the coordinate transformation \eqref{cta} through
the special (conformal) symmetries that the $z=2$ Schr\"odinger algebra
and metric possess, we can apply it to the $\mathsf{Sch}_z$ metric for
any $z$. If one does that, one finds the metric
\be
\label{gmz}
\mathsf{Sch}_z: ds^2 = -(\cos^2\omega T)^{z-2}\frac{dT^2}{R^{2z}} + 
\frac{1}{R^2}(-2dT dV - {\omega}^2(R^2 + \vec{X}^2)dT^2 + dR^2 +
d\vec{X}^2)\;\;.
\ee
Note that under this transformation the $r^{-2z}$ and $r^{-2}$ terms
of the Poincar\'e metric \eqref{metrics} do not mix and transform
separately into the corresponding terms in the metric \eqref{gmz}.

Evidently this reduces to \eqref{gma} for $z=2$. 
The $\mathsf{Sch}_z$ 
Poincar\'e metric reduces to the AdS metric either for $z=1$ or if we set 
the coefficient of the $r^{-2z}$-term to zero.  Thus by the
same token, choosing $z=1$ or setting the coefficient of the first term to zero 
in \eqref{gmz} (the two choices are related by a simple shift of $V$), we obtain a
1-parameter family of AdS metrics, namely
\be
\label{newads}
\mathsf{AdS}:\;
ds^2 = \frac{1}{R^2}(-2dT dV - {\omega}^2(R^2 + \vec{X}^2)dT^2 + dR^2 +
d\vec{X}^2)\;\;.
\ee
This is AdS in trapping coordinates, and that 
this AdS plane wave is indeed just pure AdS in disguise was already noted in
\cite{cgr}.\footnote{To see the relation between \eqref{ct1} and
the coordinate transformation given in \cite{cgr}, note that the Schwarzian derivative
$\{\tan t,t\}=2$ is constant.} Insight into this equivalence is provided by the
observation that the coordinate transformation \eqref{cta} is such that 
\be
-2dtd\xi + d\vec{x}^2 = \left(\cos^2 {\omega} T\right)^{-1}\left(
-2dT dV - {\omega}^2 \vec{X}^2 dT^2 + d\vec{X}^2\right)\;\;,
\ee
thus exhibiting the conformal flatness of the isotropic plane wave metric 
appearing on the right hand side. 
Translated into
plane wave lightcone Hamiltonians, it thus
conformally relates free particle (untrapped) and
isotropic harmonic oscillator (trapped) dynamics.
Lifting this transformation to the AdS Poincar\'e-patch by
adding the $r$-transformation in \eqref{cta}
then provides a direct means of establishing \eqref{newads},
while the reasoning of section 3.1
provides the additional insight that this coordinate system
diagonalises the action
of the modified lightcone Hamiltonian $P_+ + {\omega}^2K_-$ and the lightcone momentum $P_-$.

The metric \eqref{newads} captures the $R\ra\infty$ (bulk) behaviour of the global
$\mathsf{Sch}_{z=2}$ metric \eqref{gma}, and, in spite of its
similarity to the Poincar\'e metric, for ${\omega}>0$ this is a geodesically
complete form of the AdS metric, the trapping harmonic oscillator
potential preventing, as above, the geodesics from running off to
infinity for finite values of the affine parameter. 


Let us conclude this section with some comments on the metric
\eqref{gmz} for $z>2$. Recall from section 2 that also for $z>2$
the Poincar\'e metric is incomplete but non-singular as $r\ra\infty$.
It remains to be seen if \eqref{gmz} provides a geodesically
complete form of the metric also in this case. A new (and perhaps
at first disturbing) feature of \eqref{gmz} for $z\neq 2$ is its
$T$-dependence. However this is an unavoidable feature of global
coordinates for $z>2$. Indeed, in section 3.1 we established
the result that the Schr\"odinger space-times have no global timelike
Killing vector fields. Conversely this implies, just as in the case
of de Sitter space, that the metric in global coordinates will
necessarily be time-dependent. While this argument does not prove
that \eqref{gmz} provides a geodesically complete coordinate system
for the $z>2$ metrics, it shows that the time-dependence of \eqref{gmz}
is no reason to dismiss it. It may in any case be worth understanding
if the dependence of the metric on global time is reflected in some
manner in non-relativistic scale-invariant field theories
with $z>2$ Schr\"odinger symmetry.

\section{Scalar Fields in Global Coordinates}

In order to further study the effect of the trapping term $\omega^2(\vec{X}^2 + R^2)$
of the global metric \eqref{gma}, 
we will look at scalar fields in this section and
compare with what is know about scalars fields in the Poincar\'e patch
\cite{son,balamac}, as well as with the analysis of \cite{bafu}.

Thus, consider the Klein--Gordon equation for a massive complex scalar
field $\phi$ of mass $m_0$, 
\begin{equation}
 \frac{1}{\sqrt{-g}}\partial_{\mu}\left(\sqrt{-g}g^{\mu\nu}\partial_\nu\Phi\right)-
m_0^2\Phi=0\,, 
\end{equation} in the global $\mathsf{Sch}_{z=2}$ metric \eqref{gma}.  
We will consider modes $\Phi_{E,m}$ with a definite
(trapping) energy $E>0$ and mass (particle number) $m>0$, i.e.\
eigenfunctions of $H+a^2C=\partial_T$ and $P_-= \del_V$ of the form
\begin{equation}
\label{eq:modes}
 \Phi_{E,m}(R,\vec{X},T,V)=\ex{-iET}\ex{-im V}\phi(R,\vec X)\,,
\end{equation} 
Introducing spherical coordinates $\{\rho,\text{angles}\}$
in the $\vec X$-plane, schematically expanding $\phi(R,\vec{X}) =\sum Y_L \varphi_L(\rho)\phi_L(R)$
using spherical harmonics $Y_L$ (eigenfunctions of the Laplacian
on $S^{d-1}$ with eigenvalue $-L(L+d-2)$), one finds that 
the solutions to the separated $\rho$-equation
that are regular at the origin and well-behaved for $\rho\ra\infty$
are given by  the functions 
\begin{equation}
\label{rholag}
 \varphi_{L,n}(\rho)=\ex{-\tfrac{1}{2}\omega
 m\rho^2}\rho^LL_n^{L-1+d/2}(\omega m\rho^2)\;\;,
\end{equation} 
where the $L_n^{L-1+d/2}$ are generalised Laguerre polynomials.

The resulting $\phi_{L,n}(R)$-equation 
\begin{equation}
\label{dphi}
\phi''_{L,n}-\frac{d+1}{R}\phi'_{L,n}+\left(2Em -4m\omega(n+\frac{L}{2} + \frac{d}{4}) - \omega^2 m^2 R^2 -
\frac{m^2 + m_0^2}{R^2}\right) \phi_{L,n} = 0
\end{equation}
can then  be reduced to a standard confluent hypergeometric 
differential equation
\be
u F''(u)  + ( 1 + \frac{\Delta_+ - \Delta_-}{2} - u) F' - 
(n + \frac{L}{2} + \frac{d}{4} -\frac{E}{2\omega})F = 0
\ee
with the ansatz
\be
\phi_{L,n}(R) = \ex{-\frac{1}{2}u} u^{\Delta_+/2}F(u) \;\;,
\ee
where $u = \omega m R^2$ and 
\be
\Delta_{\pm}= \frac{d+2}{2} \pm \frac{1}{2}\sqrt{(d+2)^2 + 4 (m^2 + m_0^2)}\;\;,
\ee
The leading asymptotic behaviour of the two linearly independent
solutions $\phi^\pm$ as $R\ra\infty$ is 
\be 
\phi_{L,n}^\pm \sim \ex{\pm \tfrac{1}{2}\omega m R^2}\;\;.
\ee 
This is the analogue of the
behaviour of the Bessel functions $I_{\nu}$ and $K_{\nu}$ encountered
in the standard AdS/CFT correspondence and in the $z=2$ Poincar\'e-patch
analysis of \cite{son,balamac}.

The leading behavior of the solution near $R=0$ can also be deduced from the exact solution,
but can more readily be read off directly from 
\eqref{dphi} by neglecting the constant terms 
and, in particular, the trapping term $\omega^2 m^2 R^2$. 
The behaviour of the solutions, 
\begin{equation}\label{eq:modesnearboundary}
 \phi_{L,n}\sim R^{\Delta_\pm}\,,
\end{equation}
thus necessarily becomes identical to that found in \cite{son,balamac}
in terms of plane wave Fourier modes $\phi_{\vec{k}}$, namely
\begin{equation}
 \phi_{\vec k}\sim r^{\Delta_\pm}\,.
\end{equation}
We see that, as in the case of geodesics, the harmonic trapping term has little influence
on the dynamics near $R=0$ but strongly modifies the behaviour as $R\ra\infty$. In particular, 
the solution associated with $\phi^-$ has the characteristic harmonic oscillator fall-off behaviour
\be
\Phi_{E,m}^- \sim \ex{-\tfrac{1}{2}\omega m(R^2 + \vec{X}^2)}\;\;.
\ee
To compare with the Hamiltonian analysis of \cite{bafu}, we just note that separating out the $V$-dependence
as in \eqref{eq:modes} turns the Klein-Gordon equation into a Schr\"odinger equation and that
after the unitary transformations that eliminates the first-order derivatives from 
\eqref{dphi} and its counterpart for $\varphi_{L,n}(\rho)$
the corresponding Hamiltonian agrees with the radial Hamiltonians written down in \cite{bafu}.

\subsection*{Acknowledgements}

This work has been supported by the Swiss National Science Foundation and
the ``Innovations- und Kooperationsprojekt C-13'' of the Schweizerische
Universit\"atskonferenz SUK/CUS. We are grateful to Patrick Meessen and
Martin O'Loughlin for useful discussions.

\rnc{\Large}{\normalsize}


\begin{thebibliography}{00}
\frenchspacing
\begin{small}
\addtolength{\itemsep}{-4pt}


\bibitem{son}
  D.~T.~Son,
  \textit{Toward an AdS/cold atoms correspondence: a geometric realization of the Schroedinger symmetry},
  Phys.\ Rev.\  D78 (2008) 046003, \texttt{arXiv:0804.3972 [hep-th]}.

\bibitem{balamac}
  K.~Balasubramanian and J.~McGreevy,
  \textit{Gravity duals for non-relativistic CFTs},
  Phys.\ Rev.\ Lett.\  101 (2008) 061601, \texttt{arXiv:0804.4053 [hep-th]}.

\bibitem{lif}
  S.~Kachru, X.~Liu and M.~Mulligan,
  \textit{Gravity Duals of Lifshitz-like Fixed Points},
  Phys.\ Rev.\ D.\ 78 (2008) 106005, \texttt{arXiv:0808.1725 [hep-th]}.

\bibitem{gold}
  W.~D.~Goldberger,
  \textit{AdS/CFT duality for non-relativistic field theory},
  JHEP 0903 (2009) 069, \texttt{arXiv:0806.2867 [hep-th]}.

\bibitem{bafu}
  J.~L.~F.~Barbon and C.~A.~Fuertes,
  \textit{On the spectrum of nonrelativistic AdS/CFT},
  JHEP 0809 (2008) 030, \texttt{arXiv:0806.3244 [hep-th]}.

\bibitem{hrr}
  C.~P.~Herzog, M.~Rangamani and S.~F.~Ross,
  \textit{Heating up Galilean holography},
  JHEP 0811 (2008) 080, \texttt{arXiv:0807.1099 [hep-th]}.

\bibitem{mmt}
  J.~Maldacena, D.~Martelli and Y.~Tachikawa,
  \textit{Comments on 
string theory backgrounds with non-relativistic conformal symmetry},
  JHEP 0810 (2008) 072, \texttt{arXiv:0807.1100 [hep-th]}.

\bibitem{abmac}
  A.~Adams, K.~Balasubramanian and J.~McGreevy,
  \textit{Hot Spacetimes for Cold Atoms},
  JHEP 0811 (2008) 059, \texttt{arXiv:0807.1111 [hep-th]}.

\bibitem{duval}
  C.~Duval, M.~Hassaine and P.~A.~Horvathy,
  \textit{The geometry of Schr\"odinger symmetry in gravity background/non-relativistic CFT},
  Annals Phys.\ 324 (2009) 1158, \texttt{arXiv:0809.3128 [hep-th]}.

\bibitem{hy}
  S.~A.~Hartnoll and K.~Yoshida,
  \textit{Families of IIB duals for nonrelativistic CFTs},
  JHEP 0812 (2008) 071, \texttt{arXiv:0810.0298 [hep-th]}.

\bibitem{luca}
  L.~Mazzucato, Y.~Oz and S.~Theisen,
  \textit{Non-relativistic Branes}, JHEP 0904 (2009) 073, 
  \texttt{arXiv:0810.3673 [hep-th]}.

\bibitem{Akhavan:2008ep}
  A.~Akhavan, M.~Alishahiha, A.~Davody and A.~Vahedi,
\textit{Non-relativistic CFT and Semi-classical Strings},
  JHEP 0903 (2009) 053, \texttt{arXiv:0811.3067 [hep-th]}.

\bibitem{amsv}
  A.~Adams, A.~Maloney, A.~Sinha and S.~E.~Vazquez,
  \textit{1/N Effects in Non-Relativistic Gauge-Gravity Duality},
  JHEP 0903 (2009) 097, \texttt{arXiv:0812.0166 [hep-th]}.

\bibitem{marika}
  M.~Taylor,
  \textit{Non-relativistic holography},
  \texttt{arXiv:0812.0530 [hep-th]}.

\bibitem{Alishahiha:2009nm}
  M.~Alishahiha, R.~Fareghbal, A.~E.~Mosaffa and S.~Rouhani,
  \textit{Asymptotic symmetry of geometries with Schrodinger isometry},
  \texttt{arXiv:0902.3916 [hep-th]}.

\bibitem{sakura}
  S.~Schafer-Nameki, M.~Yamazaki and K.~Yoshida,
  \textit{Coset Construction for Duals of Non-relativistic CFTs},
  \texttt{arXiv:0903.4245 [hep-th]}.

\bibitem{hartnoll}
  S.~A.~Hartnoll,
  \textit{Lectures on holographic methods for condensed matter physics},
  \texttt{arXiv:0903.3246 [hep-th]}.

\bibitem{bkl}
  V.~Balasubramanian, P.~Kraus and A.~E.~Lawrence,
  \textit{Bulk vs. boundary dynamics in anti-de Sitter spacetime},
  Phys.\ Rev.\  D 59 (1999) 046003, 
  \texttt{arXiv:hep-th/9805171}.

\bibitem{bklt}
  V.~Balasubramanian, P.~Kraus, A.~E.~Lawrence and S.~P.~Trivedi,
  \textit{Holographic probes of anti-de Sitter space-times},
  Phys.\ Rev.\  D 59 (1999) 104021,
  \texttt{arXiv:hep-th/9808017}.

\bibitem{marolf}
  D.~Marolf,
  \textit{States and boundary terms: Subtleties of Lorentzian AdS/CFT},
  JHEP 0505 (2005) 042,
  \texttt{arXiv:hep-th/0412032}.

\bibitem{nison}
Y. Nishida, D. Son, \textit{Nonrelativistic conformal field theories},
Phys.\ Rev.\ D 76 (2007) 086004, \texttt{arXiv:0706.3746 [hep-th]}.

\bibitem{podolski} J. Podolski, \textit{Interpretation of the Siklos solutions as
exact gravitational waves in the anti-de Sitter universe},
Class.\ Quant.\ Grav.\ 15 (1998) 719, \texttt{arXiv:gr-qc/9801052}.

\bibitem{cgr}
  D.~Brecher, A.~Chamblin and H.~S.~Reall,
  \textit{AdS/CFT in the infinite momentum frame},
  Nucl.\ Phys.\  B 607 (2001) 155,
  \texttt{arXiv:hep-th/0012076}.

\bibitem{AbramowitzStegun}
  M.~Abramowitz and I.~A.~Stegun,
  \textit{Handbook of Mathematical Functions}, Dover, New York (1974).



\end{small}
\end{thebibliography}
\end{document}